\documentclass{article}

\usepackage{adjustbox}

\usepackage{arxiv}
\usepackage{amsmath}
\usepackage[utf8]{inputenc} 
\usepackage[T1]{fontenc}    
\usepackage{hyperref}       
\usepackage{url}            
\usepackage{booktabs}       
\usepackage{amsfonts}       
\usepackage{nicefrac}       
\usepackage{microtype}      
\usepackage{lipsum}		
\usepackage{graphicx}
\usepackage{natbib}
\usepackage{doi}
\usepackage{comment}
\usepackage{siunitx}

\title{High-speed imaging and coumarin dosimetry of horn type ultrasonic reactors: influence of probe diameter and amplitude}

\author{ \href{https://orcid.org/0000-0002-1219-3263}{\includegraphics[scale=0.06]{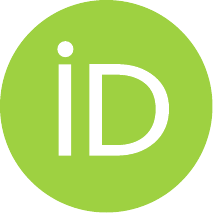}\hspace{1mm}Gianmaria ~Viciconte}\\
        Clean Energy Research  Platform\\
  Department of Mechanical Engineering\\
	King Abdullah University of Science and Technology\\
	Thuwal 23955, Saudi Arabia \\
	\texttt{gianmaria.viciconte@kaust.edu.sa} \\
     \And
 \href{https://orcid.org/0000-0001-7736-2227}{\includegraphics[scale=0.06]{orcid.pdf}\hspace{1mm}Varaha P. ~Sarvothaman}\\
       Clean Energy Research  Platform\\
  Department of Mechanical Engineering\\
	King Abdullah University of Science and Technology\\
	Thuwal 23955, Saudi Arabia \\
	\texttt{varahaprasad.sarvothaman@kaust.edu.sa } \\
 \And
 \href{https://orcid.org/0000-0002-9805-9291}{\includegraphics[scale=0.06]{orcid.pdf}\hspace{1mm}Paolo ~Guida} \\
       Clean Energy Research  Platform\\
  Department of Mechanical Engineering\\
	King Abdullah University of Science and Technology\\
	Thuwal 23955, Saudi Arabia \\
	\texttt{paolo.guida@kaust.edu.sa} \\
\And
 \href{https://orcid.org/0000-0003-1613-6052}{\includegraphics[scale=0.06]{orcid.pdf}\hspace{1mm}Tadd T. ~Truscott} \\
        Splash Lab\\
	Department of Mechanical Engineering\\
	King Abdullah University of Science and Technology\\
	Thuwal 23955, Saudi Arabia \\
	\texttt{tadd.truscott@kaust.edu.sa} \\
 \And
 \href{https://orcid.org/0000-0003-1999-2831}{\includegraphics[scale=0.06]{orcid.pdf}\hspace{1mm}William L. ~Roberts} \\
       Clean Energy Research  Platform\\
  Department of Mechanical Engineering\\
	King Abdullah University of Science and Technology\\
	Thuwal 23955, Saudi Arabia \\
	\texttt{william.roberts@kaust.edu.sa} \\ 
}

\hypersetup{
pdftitle={Towards a general description of the cavitation threshold in ultrasonic systems},
pdfsubject={q-bio.NC, q-bio.QM},
pdfauthor={Gianmaria Viciconte, Paolo Guida, Tadd Truscott, William L. Roberts},
pdfkeywords={Ultrasonicallly-Induced Cavitation, Cavitation Threshold, Acoustic Analogy },
}

\begin{document}

\maketitle

\begin{abstract}
Ultrasound driven cavitation is widely used to intensify lab and industrial-scale processes. Various studies and experiments demonstrate that the acoustic energy, dissipated through the bubbles collapse, leads to intense physicochemical effects in the processed liquid. 
A better understanding of these phenomena is crucial for the optimization of ultrasonic reactors, and their scale-up. In the current literature, the visual characterization of the reactor reactivity is mainly carried out with sonoluminescence and sonochemiluminescence. These techniques have limitations in the time resolution since a high camera exposure time is required.
In this research, we proposed an alternative method, based on coumarin dosimetry to monitor the hydroxylation activity, and high-speed imaging for the visualization of the vapor field.
By this approach, we aim to capture the structure and the dynamics of the vapor field, and to correlate this with the chemical effects induced in the ultrasonic reactor. This characterization was carried out for four different ultrasonic probe diameters (3, 7, 14 and 40 mm), displacement amplitudes and processing volumes.
Key findings indicate that the probe diameter strongly affects the structure of the vapor field and the chemical effectiveness of the system.
The proposed methodology could be applied to characterize other types of ultrasonic reactors with different operating and processing conditions.
\end{abstract}


\section{Introduction}
Ultrasound finds a rather unique position in human lives with its utility ranging from human diagnostics to intensifying chemical reactions, and aiding in detection of pipe corrosion. The passage of sufficiently intense ultrasound waves, into a liquid medium, generates the phenomenon of cavitation, with varying extent of physical and chemical transformations brought about in the bulk of the liquid. Broadly 20 to 40 kHz is termed as conventional power ultrasound used for physical and biological applications, with a great utility to process intensification. 40 to 500 kHz can be termed as the range for observing sonochemistry, with applications in sonolysis-based water treatment and food processing. 

The interest in a systematic understanding of ultrasound reactors has been the subject of research from the late 1990's, propelled by numerous experimental reports through the 1900's \citep{richards1927chemical, chen1967effect, okada1972effect}. Due to this profound interest, systematic efforts in expanding the depth of investigation of ultrasound based processes emerged experimentally \citep{ashokkumar1999ultrasound, suslick1999applications, mason1992industrial, gogate2002mapping, gogate2004sonophotocatalytic, gogate2008cavitational, asgharzadehahmadi2016sonochemical, yao2020power} and numerically \citep{dahnke1997numerical, laborde2000fluid, hussain2016numerical, johansson2017design, yasui2021numerical,  ferkous2023sonochemical, sboev2024numerical}, and some to correlate them \citep{mason2011new, abid2013effect, jambrak2014effect, fu2020sono}. This past research was crucial for a comprehensive characterization of the sonochemical reactors which can lead to the design of optimized reactors. The key aspect is the identification of the phenomenological relation between the structure of the acoustic pressure field, the cavitation activity in the domain, and the physio-chemical effects induced by the dynamics and collapse of the vapor bubbles and clusters \citep{suslick1990sonochemistry, didenko1999hot, wagterveld2011visualization}.

The efficiency of these types of reactors has to be evaluated by relating the physical or chemical effects, for which the reactor was designed, with the acoustic energy introduced into the system by the ultrasound source.
Some of the most relevant physical effects induced by the collapse of the cavitation bubbles are: emulsification \citep{canselier2002ultrasound,cucheval2008study}, sonofragmentation \citep{sander2014sonocrystallization, kim2018effects} and leaching \citep{luque2003ultrasound, narayana1997leaching}. One of the most advanced tools for the stand-alone investigation of these phenomena is the high-speed imaging, since it is able to provide extremely fine spatial and temporal resolutions \citep{cucheval2008study, wagterveld2011visualization, bovcek2023dynamics, perdih2019revision}.

The chemical effects of ultrasound cavitation, in an aqueous solution, comprise the generation of different radical species such as OH$\cdot$, OOH$\cdot$, and so on. These are due to the dissociation of water molecules trapped in the cavitation bubble, during their collapse. The radicals generated could play a significant role in increasing the rate of different chemical reactions based on the solute processed \citep{suslick1997chemistry, sarvothaman2023cavitation}. The most used methodology to quantify the radicals, generated by sonochemistry, is the "chemical dosimetry" \citep{iida2005sonochemistry, martinez2010salicylic}. Different techniques are available and they allow a quantitative estimation of the radicals, through ex-situ analysis of the processed samples.
Two visual techniques are commonly used for in-situ characterizations of the reactive zones of sonochemical reactors: sonoluminescence (SL) and sonochemiluminescence (SCL). In these techniques, a camera sensor collects the light emission from the collapse of cavitation bubbles (SL) or from the reaction between hydroxyl radicals (OH$\cdot$) with luminol, deliberately dosed to the system (SCL) \citep{rivas2012sonoluminescence}. Qualitative information can be obtained to compare the reactive response/performance between different configurations, operational modes, or reactors. Regarding these two techniques, it is important to highlight that an extremely high exposure time (15 seconds or longer) is required to obtain a detectable signal on the camera sensor \citep{rivas2012sonoluminescence, bampouli2023understanding, bampouli2024importance,pflieger2015effect}. This constitutes a significant disadvantage, since the information is averaged over a time span which is order of magnitudes higher than the characteristic time of the ultrasonic waves and cavitation phenomena.

Considering the shortcomings of SL and SCL, a two-pronged approach of high-speed imaging of the system and an independent chemical estimation could be of better utility to individually quantify the physical and chemical effects of the cavitation system thereof.
In the present work, this was achieved by monitoring the hydroxylation product of coumarin and correlate this with the structure and the dynamics of the cavitation vapor field, visualized by high-speed imaging. The high-speed visualization experiments were carried out at a frame rate relevant to capture the dynamics and the structure of the vapor field, generated by cavitation in the liquid domain.

Although ultrasound reactors can be realized in a variety of different configurations (e.g. horn type, transducer or ultrasonic bath) and operational modes (e.g. working frequency, area of the emitting surface, acoustic power etc.) \citep{adamou2024ultrasonic, asgharzadehahmadi2016sonochemical}, the present case study concerns a lab-scale, horn-type ultrasound device with a working frequency of 24 kHz as it is a type of device commonly used, provides a direct contact of the emitting surface with the liquid medium, and this frequency has the potential to be applied to continuous flow processing \cite{zhu2024acoustic, lakshmi2023acoustic, malek2020ultrasonication, chuah2022review}. Several studies have investigated the impact of the design parameters on the efficiency of sonochemical reactors \citep{wood2017parametric,bampouli2023understanding}. 

For the horn-type configuration, the parameters of primary importance are: probe diameter, vibration amplitude, and volume processed \citep{bampouli2024importance}, these are investigated here. The trends obtained in terms of quantitative hydroxylation will help ascertain the behavior of the vapor structure contributing to effective chemical activity. This approach would allow us to highlight the working principles of the sonochemical reactor and understand the main scaling parameters. Although the analysis has been carried out on a single frequency ultrasonic device, the intent is to propose a methodology that could be applied, with some variations, to study other types of reactors and transducers, characterized by different operating conditions and process objectives.

\section{Materials and Methods}
\subsection{Experimental setup}
For the generation of the ultrasonic waves in the liquid medium, a Hielscher device UP400S (24 kHz) was used. This device absorbs a maximum power of \SI{400}{W} and titanium probes having different diameters can be installed on it. The vibration amplitude of the probe can be regulated, on the ultrasonic device, in terms of percentage with respect to the maximum (from 30\% to 100\%). At every percentage amplitude, it is possible to associate the actual amplitude in micrometers and the calorimetric power (relevant parameters are given in Tab. \ref{Tab1}, details are presented in sections 2.4 and 3.1).
In the present study four different probe diameters ($d$) have been tested: 3, 7, 14 and 40 mm. For the coumarin dosimetry experiments (details in Section 2.3), the water solution was placed in a cylindrical glass beaker. Two beaker sizes were used to process two different solution volumes ($V$): 200 and 400 mL. The shape of the container was chosen to have the wall container several diameters away from the vibrating surface of the probe.    
For every configuration, the immersion depth of the probe in the solution was kept constant to \SI{25}{mm}. The temperature was monitored with a thermocouple immersed in the liquid at the corner of the beaker, in the form of a wire, so as to avoid any alterations to the ultrasound cavitation field. The electrical power, absorbed by the ultrasound device, was measured with a socket type multimeter. 
\subsection{Chemicals used and analytical techniques employed}

The concentration of radicals, produced in an aqueous solution, can be measured with dosimetry techniques. These rely on the reaction between an initial reactant (solute) as a scavenger for the radicals formed during the cavitation activity. In other words, the solute by partially reacting with radicals is converted into detectable products. A lot of focus is placed on hydroxyl radicals as these have the second highest oxidation potential ($E^0$=2.8 eV) only inferior to that of Fluorine. 
In this work, the so-called coumarin dosimetry was used to quantify the hydroxyl radicals. In the presence of radicals generated by cavitation activity, coumarin molecules react to form different species such as 3-hydroxycoumarin, 4-hydroxycoumarin, 5-hydroxycoumarin, 6-hydroxycoumarin, 7-hydroxycoumarin, and 8-hydroxycoumarin and further reaction of these lead to formation of di-hydroxycoumarin molecules as well \citep{louit2005reaction, nopel2023experimental}. 
The reaction chemistry of coumarin was not of primary interest in this study, rather the interest was to employ coumarin as a "radical trap" to quantify cavitation effects. Other methods such as the Weissler reaction \citep{morison2009limitations} and dosimetry with salicylic and terephtalic acids are available, however, they have their limitation either in terms of not being specific to the cavitation activity or the need for operating at an altered pH of the system respectively \citep{martinez2010salicylic, arrojo2007application, iida2005sonochemistry}. 

It is important to clarify that with the proposed dosimetry technique, just a small part of the total hydroxyl radicals, produced into the system, is detected. Indeed, a consistent part of radicals, after being generated, are reabsorbed by recombination reactions. This is an intrinsic drawback of every dosimetry technique: it is impossible to measure all the radicals produced in the system since they are extremely reactive and ephemeral \citep{iida2005sonochemistry, martinez2010salicylic}.

Furthermore, some radicals react with the chemical probe by taking alternative reaction pathways and forming products which are not easy to monitor \citep{louit2005reaction, nopel2023experimental}. Regarding this aspect, some authors have attempted to estimate the total hydroxyl radicals. For example, in the context of coumarin dosimetry, Zhang et al. \citep{zhang2013quantitative} estimated that just 6.1\% of all the OH$\cdot$ radicals produced are scavenged as 7-hydroxycoumarin (7OHC). However, as suggested by De-Nasri et al. \citep{de2021quantification}, this approximation may not be valid and consistent for other advanced oxidation processes, since with the probe based methods of OH$\cdot$ quantification, the hydroxylated product yields are process specific.

Considering the complexity of the chemical problem, the present study does not aim at a deeper investigation of all the different chemical pathways of the radicals generated.
Coumarin dosimetry is here used just as a method to characterize the sonochemical reactor. For these reasons, various chemical parameters such as operating pH or initial temperature have not been investigated.
The assumption made is that the amount of OH$\cdot$ estimated, by monitoring the fluorescent product (7OHC), is representative of the chemical efficiency of the configurations tested. This assumption has been used in previous cavitation based investigations \citep{sarvothaman2024evaluating, de2021quantification,de2022quantifying}.

The 7OHC, in an aqueous solution, can be detected by fluorescence spectroscopy, since the fluorescence spectrum responds to the 7OHC (emission wavelength: 332 nm, excitation wavelength: 450 nm). For a quantitative estimation, a calibration curve was prepared across a range of anticipated 7-hydroxycoumarin concentrations. This curve, where the concentration of 7OHC is a function of the intensity of the fluorescence signal, was used to identify the concentration of the compound in samples collected from the experiments. 

The chemicals used for the experimental campaign, coumarin ($\geq$ 99\%, CAS number: 91–64-5) and 7-hydroxycoumarin (99\%, CAS number: 93–35-6), were procured from Sigma-Aldrich. A stock solution of 1 g/L of coumarin in DI water was prepared, to ensure complete coumarin dissolution. Furthermore, a stock solution of 50 mg/L of 7-hydroxycoumarin was created using DI water, for the preparation of the samples, used to generate the calibration curve. The 7OHC compound can be monitored, on the fluorometer (Cary Eclipse Fluorescence Spectrometer), with an excitation wavelength of 332 nm and an emission wavelength of 450 nm.

To be consistent with other works present in literature \citep{de2022quantifying, sarvothaman2024evaluating}, for all the experiments, the initial concentration of coumarin was kept at 15 mg/L. This low concentration was also adopted to not affect the cavitation activity \citep{zupanc2020anomalies}. Thus validating the assumption that the cavitation field in pure water is representative of the cavitation activity during the coumarin experiments. The dosimetry experiments were carried out for a maximum of \SI{28}{min}, without actively cooling down the system. To minimize the evaporation of the processed solution, the headspace between the ultrasonic probe and the beaker was covered using an aluminum foil. The evaporation was estimated with a mass balance at the end of every experimental run. Based on the amount of water evaporated, the measured concentrations were corrected. All the experiments have been carried out with an initial temperature of the solution around 23 ± 1 $^\circ$C. Over the processing time, at every 4 minutes, a 4 mL volume was sampled from the reactor by using a calibrated pipette. The samples collected were analyzed with the fluorometer to quantify the concentration. The uncertainty of the measurements was estimated by repeating the fluorometer analysis three times on the same sample. It was calculated to be within ± 2\%.
\subsection{High-speed imaging}

The cavitation vapor field, induced by ultrasonic waves at 24 kHz, can be observed with a high-speed camera and the backlighting technique, where the scene is placed between the light source and the camera. Due to the refraction of light at the interface between the phases, it is possible to visualize shadows of objects and the spatial and temporal evolution of the vapor field within the domain \citep{vznidarvcivc2014attached, kozmus2022characterization}.
A schematic illustration of the setup employed is shown in a previous study of the authors \citep{viciconte2023towards}. A LED light source (Godox SL-200W II), emitting white lite non-coherently, was placed on the back side of a water container and aligned with the axis of the probe. The scene was captured using the high-speed camera Phantom v2511 placed in front of the container, at the opposite side of the LED light source. A frame rate of 200,000 fps (200 kHz) at 256x256 pixels was used to capture the phenomenon. The camera was equipped with a 105 mm Nikon MICRO Lens, providing a resolution of 0.055 mm/pixel. 
The focal plane was placed on the axis of the vibrating probe, and the maximum aperture of the lens was used (f/2.8). This allows us to have a narrow depth of field, having in focus just the vapor cavities in the neighborhood of the plane of focus. A large aperture also reduces the exposure time required, as it increases the light reaching the sensor. An exposure time of \SI{1.5}{\micro s} was set. For the visualization experiments a cuboidal container, made of transparent acrylic and having dimensions of 95 mm × 190 mm × 190 mm, was partially filled with Milli-Q® water at 23°C. The depth of water in the container was 70 mm and the tip of the probe was immersed 25 mm below the free surface.
\newline

\begin{figure*}[h!]
    \centering
    \includegraphics[width=1\linewidth]{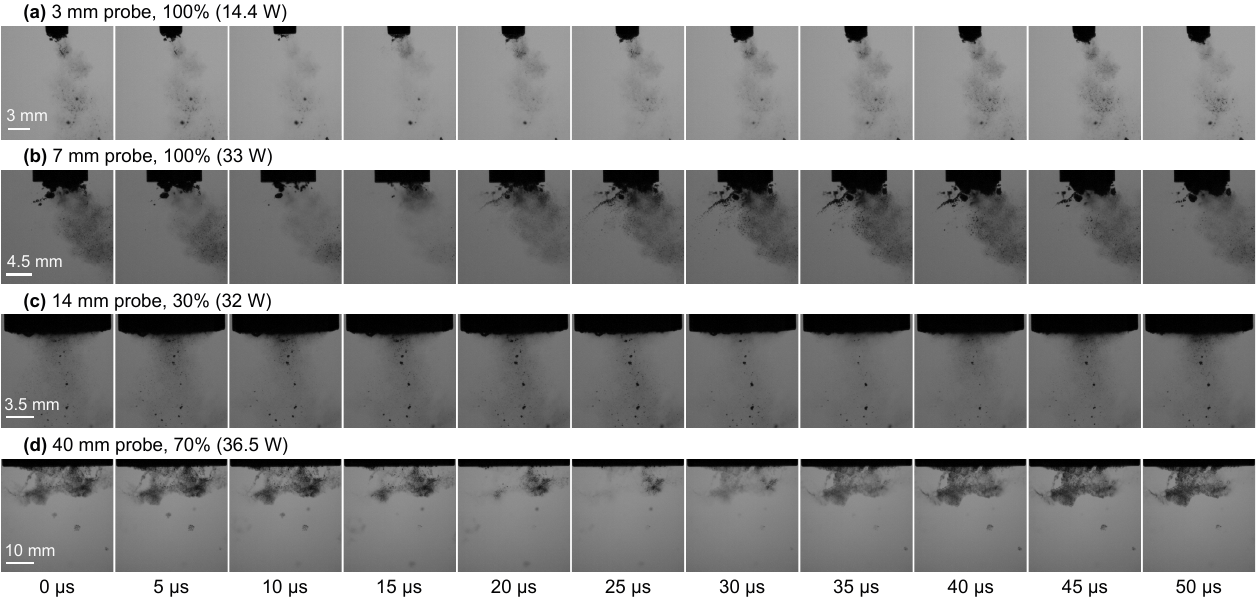}
    \caption{High-speed frame sequences for different ultrasound probe (Videos in the SI). (a-d) are individually labeled by their probe diameter, nominal amplitude and input calorimetric power (definition in Eq. \ref{equazione_2}). Time between frames is the same for (a-d) as marked.}
    \label{Fig1}
\end{figure*}

The frames, obtained with high-speed visualization, for some of the experimental configurations, are presented in Fig. \ref{Fig1}.
The shadow of the cylindrical probe and the vapor cavities are defined by the dark areas of the image, while the light region represents the liquid domain (Fig. \ref{Fig1}). 

\begin{figure*}[h!]
    \centering
    \includegraphics[width=0.5\linewidth]{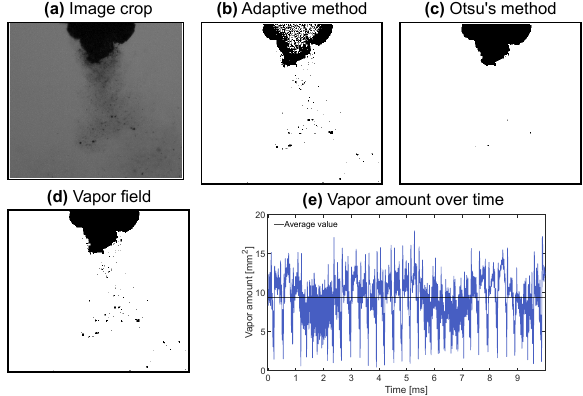}
    \caption{Image process for a 7 mm probe with a nominal amplitude of 70\%: (a) Cropped image (grayscale frame from the recorded video); (b) binarized image using the adaptive thresholding method; (c) binarized image using Otsu's method; (d) final binary image obtained using an hybrid procedure of the two methods; (e) total vapor amount over time from the high-speed image sequence. The unit measure of the y-axis is \SI{}{mm^2} since the backlighting visualization provide a 2D representation of the vapor phase.}
    \label{Fig2}
\end{figure*}

The cavitation field presents a complex morphology of vapor structures (Fig. \ref{Fig1}). In this regard, the vapor field generated by the 3, 7 and 14 mm probes, can be studied by identifying two different regions: the first one, in the near-field, where a large vapor cluster is attached to the probe tip, the second one, in the far-field, characterized by a cloud of smaller bubbles. 
On the contrary, in the configuration with the 40 mm probe, there is no evidence of the vapor cluster attached to the tip (Fig. \ref{Fig1}d). Quantitative data, on the vapor phase present in the domain, can be extracted by binarizing the original grayscale images (Fig. \ref{Fig1}). The binarization procedure allows for a clear distinction between the phases, with the liquid represented in white (pixel value = 1) and the vapor in black (pixel value = 0). Because of the structural complexity of the vapor field, the image binarization was carried out with a hybrid strategy. This is summarized, for a single frame, in Fig. \ref{Fig2}. Two different functions available on Matlab \citep{MATLAB} were used: the \textit{adaptive thresholding method} and the \textit{Otsu's method}. The first one provides a good binary representation of the vapor cloud in the far field, but it fails in the binarization of the vapor cluster attached to the tip of the probe (Fig. \ref{Fig2}b). The second one is able to preserve the shape of the cluster in the near-field, but fails with the vapor cloud in the far-field (Fig. \ref{Fig2}c). For both methods, the sensitivity parameters have been calibrated to preserve the shape and the boundaries of the vapor structures, visible in the original grayscale images (Fig. \ref{Fig2}a). A similar approach, for the study of cavitation induced by ultrasonic probes, has previously been used by other authors \citep{kozmus2022characterization,vznidarvcivc2014attached}. fter obtaining the binary representation (Fig. \ref{Fig2}d) of the high-speed sequence, the total amount of vapor, as a function of time, was computed by multiplying the number of black pixels by the effective area of a pixel, known by optical calibration. The chart in Fig. \ref{Fig2}e shows the vapor amount over time for one experimental case. The vertical solid line indicates the temporal average value (\(A_{vap}\)). The same procedure can be applied to extract other information from the binary image, like the evolution in time of the area of the vapor cluster (\(A_{clust}\)).

The dynamics of the vapor structures were analyzed with the DFT (Discrete Fourier Transform) of the pixels intensity of the original high-speed image sequence (Fig. \ref{Fig1}). The DFT can be done, through the FFT (Fast Fourier Transform) algorithm, on different portions of the image, depending on the dynamics of the structure of interest. The DFT of the image portion related to the far-field reveals one significant peak at 23.8 kHz. This means that the vapor bubbles, in that specific region, mainly oscillate at the same frequency of the ultrasound probe. Instead, the DFT of the near-field reveals, together with the peak at 23.8 kHz, the presence of other peaks in the low frequency range of the spectrum (Fig. \ref{Fig8}). This suggests that the vapor cluster oscillates at frequencies lower than the excitation one. These outcomes are important to characterize the dynamic behavior of the system.

\begin{table}[h!]

\caption{Table containing the most relevant parameters related to the experiments performed with different probe diameters and displacement amplitudes.}
\centering
\begin{tabular}{l|r|r|r|r|r|r|r|r}
\(d\) [mm] & $A_{nom}$ [\%] & $A_{r}$ [\SI{}{\mu m}] ($\pm$ 2.35)& \(\dot{W}_{el}\) [W] ($\pm$ 0.5) & \(\dot{W}_{cal}\) [W] ($\pm$ 0.5) & \(\eta_{dev}\) & \(I\) [\SI{}{W/mm^2}] & (\(A_{vap}\)) [\SI{}{mm^2}] \\\hline 
3&30&23.54&10.8&1.1&0.10& 0.16&0.28\\  
3&50&25.89&17&5.6&0.33& 0.79  &0.41 \\ 
3&70&47.08&23.8&9.1&0.38& 1.29 &0.79 \\
3&90&62.38&35.5&13.1&0.37&  1.85 &1.49 \\
3&100&64.74&39.3&14.4&0.37&  2.04 &2.19\\\hline
7&50&36.39&29.3&13.6&0.46& 0.35  &5.69 \\
7&70&50.47&42&21.5&0.51& 0.55 &   9.38\\
7&90&64.55&60.5&31.5&0.52& 0.82 &  11.18\\
7&100&68.08&65.2&32.9&0.50& 0.85 &  13.99 \\\hline
14&30&19.95&55&32&0.58& 0.21  & 4.03 \\
14&50&29.34&84&54&0.64&  0.35 & 7.27\\
14&70&39.91&121&80.8&0.67& 0.53 &  9.97 \\
14&90&53.99&187&124.5&0.67& 0.81& 17.51 \\
14&100&55.16&202&138.5&0.69&  0.90 &19.42 \\\hline
40&30&2.35&43&21.5&0.50& 0.017 &20.70\\
40&50&2.35&58.9&28&0.48& 0.022  &34.91 \\
40&70&4.70&70&36.7&0.52& 0.029 &43.91\\
40&90&5.88&95&56.2&0.59& 0.045 &47.83 \\
40&100&5.88&103&59.3&0.58& 0.047 &56.57\\
\end{tabular}
\label{Tab1}
\end{table}

The same backlighting setup used for the visualization of the cavitation field (Fig. \ref{Fig1}) was also used for measuring the actual vibration amplitude of the ultrasonic probes. This was crucial since the regulation system of the ultrasound device just provides the percentage amplitude ($A_{nom}$). Since the displacement is expected to be on the order of magnitude of micrometers, a high level of magnification is required. Thus, the Phantom v251 was combined with a Mitutoyo 10X objective lens, to provide a resolution of \SI{2.3}{\micro m}/pixel). The sequence of magnified frames obtained, for the 7 mm probe, is provided in Fig. 1 of the Supplementary Information (SI). The displacement amplitude can be visualized only by framing the corner or the lateral surface of the probe because the tip is always covered is covered by vapor cavities for all working conditions.
The displacement amplitude of the probe ($A_{r}$) was measured with a tracking procedure, implemented on Matlab \citep{MATLAB}. Starting from the frames sequence, the displacement can be tracked by selecting one column of pixels and by taking the difference between adjacent pixels. From the resulting array, the element with the maximum value defines the position of the probe tip at every time step. This tracking procedure is described more in detail in a previous work of the authors \citep{viciconte2023towards}. The values obtained, for the different configurations tested, are shown in Tab. \ref{Tab1}. The DFT, of the probe displacement over time, revealed a frequency of $f_p=23.78$ kHz (slightly different from the advertised value of the device of 24 kHz). 

\subsection{Measurement of acoustic spectra}
The acoustic signal was recorded by inserting a hydrophone (Aquarian AS-1, receiving sensitivity: 40 $\mu$V/Pa, linear range: 1 Hz to 100 kHz ±2 dB) in the liquid domain. This device works with a direct connection to an oscilloscope. For the signal acquisition, two different sampling frequencies were set: 12.5 and 31.25 MHz. Due to the oscilloscope setting, these frequencies correspond to two different time intervals acquired, respectively 16 and 6.4 ms.
The signals, measured by the hydrophone, have been used to further characterize the cavitation activity in the ultrasonic reactors, by comparing the spectra related to the different configurations.

\section{Results and discussion}
\begin{figure*}[t!]
    \centering
    \includegraphics[width=1\linewidth]{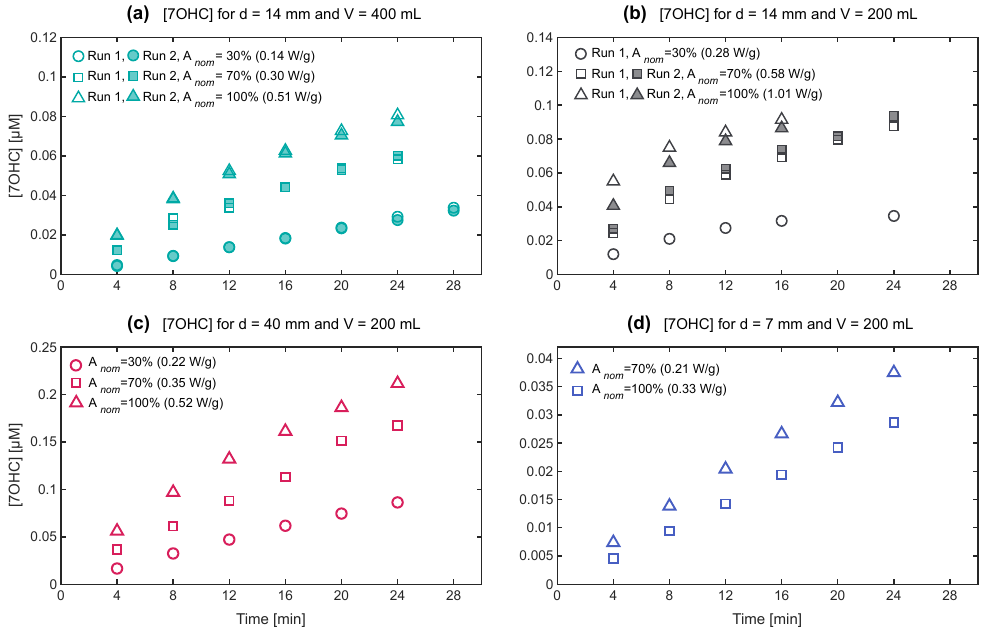}
    \caption{Concentration profile of 7-hydroxycoumarin for the experimental configurations tested as titled in (a-d). The different marker symbols indicate experiment done with different nominal amplitudes. The specific electric power (W/g) was measured with a power meter.}
    \label{Fig3}
\end{figure*}

The results of the 7OHC conversion are presented in Fig. \ref{Fig3} for different probe configurations and amplitudes.
The concentration over time for the 14 mm probe was estimated for two different solution volumes: 200 and 400 mL (Fig. \ref{Fig3}a, b). Furthermore, for these cases, a repeatability study was conducted (Run 1 and Run 2 in Fig. \ref{Fig3}a, b). As can be observed, the data reveal a good repeatability of the experiments. The probes having diameters 40 mm and 7 mm were tested only with the volume solution of 200 mL (Fig. \ref{Fig3}c, d). No data related to the 30\% amplitude for the 7 mm probe, and to the 3 mm probe are shown, due to the poor energy input and the consequent low chemical conversion for these configurations (below the detection limit of the fluorometer).

An alternative experiment was conducted to prove the reliability of the coumarin dosimetry method in producing hydroxylated products only in the presence of cavitation activity. The same initial solution, used for the parametric study, was heated by using a heating plate, equipped with a mechanical stirrer. A thermal power of 330 W was provided to the system and the experiment was stopped at 14 minutes, with a system temperature of 95$^\circ$C. The fluorometer analysis of the sample has revealed no coumarin conversion into 7OHC. This result shows no oxidation in the presence of a heat source and atmospheric air, confirming the coumarin dosimetry as an unambiguous method for the OH radicals produced by cavitation activity. 

For every configuration tested, the concentration of 7OHC increases with time and with the vibration amplitude of the ultrasonic probe (Fig. \ref{Fig3}). This is expected since an increase in the amplitude leads to an increase in the energy introduced into the system, resulting in an increase in energy dissipated due to the formation of radicals. 

It is interesting to note that the configurations, characterized by a lower specific power, exhibit a linear trend (e.g. those in Fig. \ref{Fig3}d). This is probably due to the dependency between the power dissipated by the probe and the liquid temperature. Since those configurations are characterized by a low increment of the liquid temperature (Fig. 2 in SI), the variation of the physical property (acoustic impedance and viscosity) is small. As a consequence, the power dissipated by the probe and that absorbed by the device remains almost constant (Fig. 3 in SI) during the entire duration of the process. This implies a constant rate of the radicals production through time (Fig. \ref{Fig2}). On the contrary, in the configurations characterized by a large increase in temperature (Fig. 2 in SI), the significant variation of the liquid properties impacts the power dissipated by the probe. This entails a decrease, with processing time, in the electrical power (Fig. 3 in SI) and the production rate of radicals (Fig. \ref{Fig3}).

Since the energy contributions (energy absorbed by the ultrasound device and energy introduced into the system) are different for every configuration, the results cannot be compared in the form shown in Fig. \ref{Fig3}.
\begin{figure*}[h!]
    \centering
    \includegraphics[width=1\linewidth]{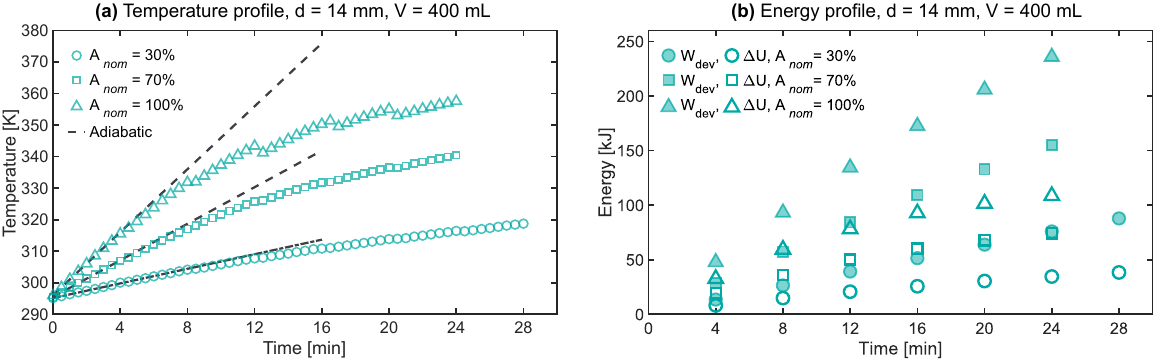}
    \caption{Temperature (a) and energy profile (b) for the configuration with the 14 mm probe diameter and a volume processed of 400 mL. (a) Evolution in time of the solution temperature, measured with a thermocouple, for three different nominal amplitudes. (b) Evolution in time of the electric energy used by the ultrasound device and the internal energy of the system. The difference between the two contributions is due to the efficiency of the device and to the heat exchanged.}
    \label{Fig4}
\end{figure*}
For this reason, to compare the conversion data related to different configurations, an energy balance of the system is required.

\subsection{Energy balance and comparison}

The energy introduced in the system, through the work done by the ultrasonic probe (\(W_{in}\)), is partly spent to increase the internal energy of the system (\(\Delta U\)) and partly exchanged, as heat, towards the environment ($Q_{out}$). Under the assumption of ideal incompressible behavior \citep{Gyf}, the energy balance can be written in the following form:

\begin{equation}
\label{equazione_1}
{W}_{in}-{Q}_{out}={\Delta U}=m \cdot {c}_{p} \cdot ({T}-{T}_{i}),
\end{equation}
where, $m$ is the mass of the liquid, $c_{p}$ is the specific heat capacity of the liquid medium (in this case water), $T$ is the temperature of the system (Fig. \ref{Fig4}a), and $T_{i}$ is the initial temperature. Formally, also the energy stored by the chemical reactions should be taken into account. However, considering the law concentrations of products that characterize these type of sonochemical reactors (Fig. \ref{Fig3}), this contribution can be reasonably neglected in the energy balance. Only during the first few minutes of the process, a further approximation can be made, by considering the system to be adiabatic. Indeed, due to the small volume of the liquid domain, the heat exchanged by the system towards environment can be neglected. This assumption is confirmed by the linear trend shown by the temperature curve in the first few minutes (dashed line in Fig. \ref{Fig4}a). Under this assumption, the energy balance simplifies and the effective power, provided by the ultrasonic probe (\(\dot{W}_{in}\)), can be expressed as:

\begin{equation}
\label{equazione_2}
\dot{W}_{in}=\dot{\Delta U}=m \cdot {c}_{p} \cdot \frac{dT}{dt}, 
\end{equation}
where \(\dot{\Delta U}\) is the rate of change of the internal energy, and \(\frac{dT}{dt}\) the time derivative of the temperature. Since the temperature was monitored with a thermocouple inserted in the solution (Fig. \ref{Fig4}a), Eq. \ref{equazione_2} allows us to estimate \(\dot{W}_{in}\), at the beginning of the experiment, for every configuration (Tab. \ref{Tab1}). The procedure mentioned constitutes the calorimetric methods (\(\dot{W}_{in}=\dot{W}_{cal}\)), usually employed to characterize sonochemical reactors \citep{bampouli2024importance, bampouli2023understanding, pflieger2015effect}. Considering the acoustic power provided by the probe equal to the calorimetric power is a reasonable assumption since most of the energy is dissipated as heat into the system.

\begin{figure*}[h!]
    \centering
    \includegraphics[width=1\linewidth]{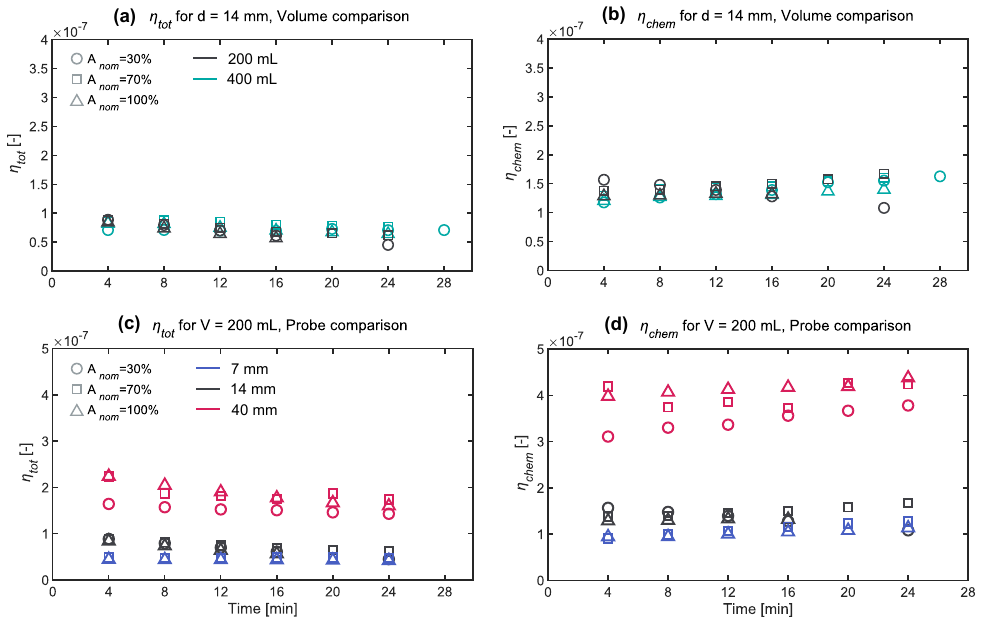}
     \caption{(a) Influence of processing volume on total efficiency over time for the 14 mm probe. (b) Influence of processing volume on chemical efficiency for the 14 mm probe [symbol legends are identical to chart (a)]. (c) Total efficiency for different probe diameters. (d) Chemical efficiency for different probe diameters [symbol legends are identical to chart (c)].}
    \label{Fig5}
\end{figure*}

The efficiency of the device ($\eta_{dev}$), at the beginning of the run, can be computed:
\begin{equation}
\label{equazione_2}
\eta_{dev}=\frac{\dot{W}_{in}}{\dot{W}_{el}},
\end{equation}
where $\dot{W}_{el}$ is the electrical power absorbed by the ultrasound device and measured with a digital power meter (Tab. \ref{Tab1}). The device efficiency, for every configuration tested, is listed in Tab. \ref{Tab1}.

In the following analysis, to normalize the concentration of 7OHC, two different energy contributions are considered: the specific variation of the internal energy ($\Delta u=\Delta U/m$) and the electrical energy, absorbed by the device, in specific terms ($w_{el}=W_{el}/m$ ). The trend over time of these two contributions is shown in Fig. \ref{Fig4}b. To obtain a non dimensional parameter, the molar concentration of 7OHC ($C_{7OHC}$) can be multiplied by the bond dissociation energy of OH ($BDE_{OH}$=497.1 kJ/mol) and divided by $\Delta u$. In this way, a sort of chemical efficiency of the system ($\eta_{chem}$) can be computed:
\begin{equation}
\eta_{chem}=\frac{C_{7OHC}\cdot BDE_{OH}}{\Delta u}.
\label{equazione_3}
\end{equation}
Since just part of the radicals generated is measured and the enthalpy of formation of 7OHC was not considered, the previous definition (Eq. \ref{equazione_3}) is formally incorrect.  
However, it provides an estimation of the portion of energy which is effectively spent for generating the radicals, comparing to the global energy that remains in the system ($\Delta u$). For this reason, Eq. \ref{equazione_3} allows us to evaluate the relative efficiency of the different configurations tested. Another non-dimensional parameter can be introduced to define a sort of total efficiency of the system:

\begin{equation}
\eta_{tot}=\frac{C_{OH}\cdot BDE_{OH}}{w_{el}}.
\label{equazione_4} 
\end{equation}
This provides an indication of the energy used for the generation of radicals, compared to the total electric energy absorbed by the ultrasound device.
The two dimensionless parameters (Eq. \ref{equazione_3}, \ref{equazione_4}) can be used for a consistent comparison of the result provided by the different configurations (Fig. \ref{Fig5}). 

It is noteworthy that the trend of the chemical efficiency (Fig. \ref{Fig5}b, d) is almost constant with the process time. This means that the coumarin conversion is just proportional to the energy dissipated by the probe and that the bulk temperature trend (different for every configuration, Fig. 2 in the SI) does not significantly modify the reaction rate and the availability of OH* radicals into the system (consequence of the cavitation activity). 
Chart (a) and (b) are respectively related to the parameters $\eta_{tot}$ and $\eta_{chem}$ as a function of the processing time, for the configurations with the 14 mm probe and two different process volume (200 and 400 mL). As observable, for both cases, the variation of amplitude and the process volume does not significantly impact the system efficiency. More interesting findings can be made by comparing the results coming from experiments with fixed process volume (200 mL) and different probe diameters (7, 14, 40 mm) (Fig. \ref{Fig5}c, d). For both the parameters, $\eta_{tot}$ (Fig. \ref{Fig5}c) and $\eta_{chem}$ (Fig. \ref{Fig5}d), only the configuration with a 40 mm probe shows sensitivity with respect to the displacement amplitude. Indeed, the efficiency for both parameters increases as the amplitude increases. Instead, the efficiency for the 7 and 14 mm probes is almost insensitive to the amplitude. In global terms, for both the parameters, the 40 mm probe shows a significantly greater efficiency than the 7 and 14 mm ones (Fig. \ref{Fig5}c, d). The explanation for this is to be found in the different structures of the vapor field generated by the probes (Fig. \ref{Fig1}).
\subsection{Vapor field, structure and dynamics}
 
\begin{figure*}[h!]
    \centering
    \includegraphics[width=0.5\linewidth]{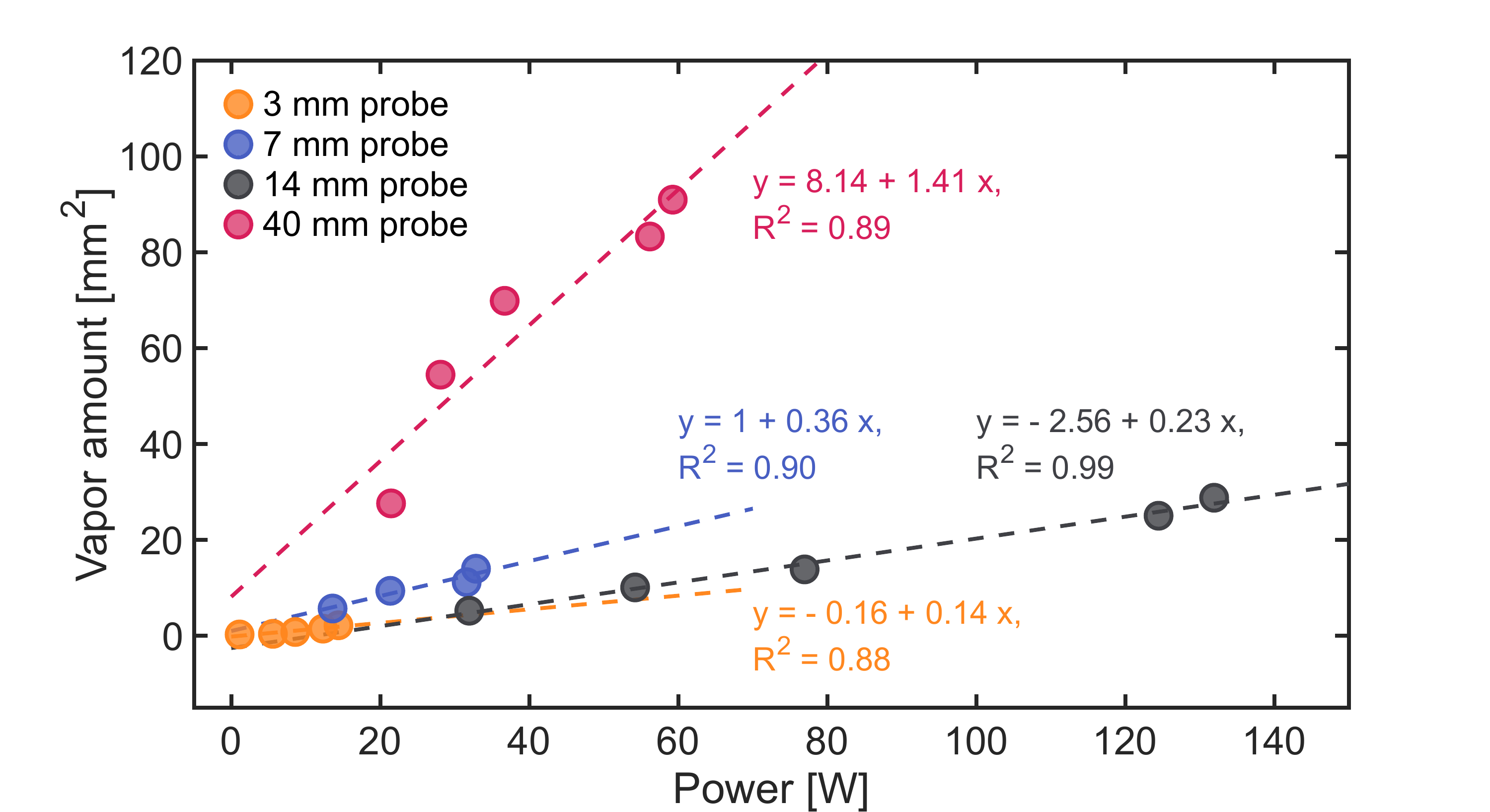}
    \caption{Average vapor amount over the calorimetric power (the vapor amount was estimated by binarizing the raw high-speed images). The dashed line are the linear regressions of the scatter data.}

    \label{Fig6}
\end{figure*}

\begin{figure*}[h!]
    \centering
    \includegraphics[width=0.5\linewidth]{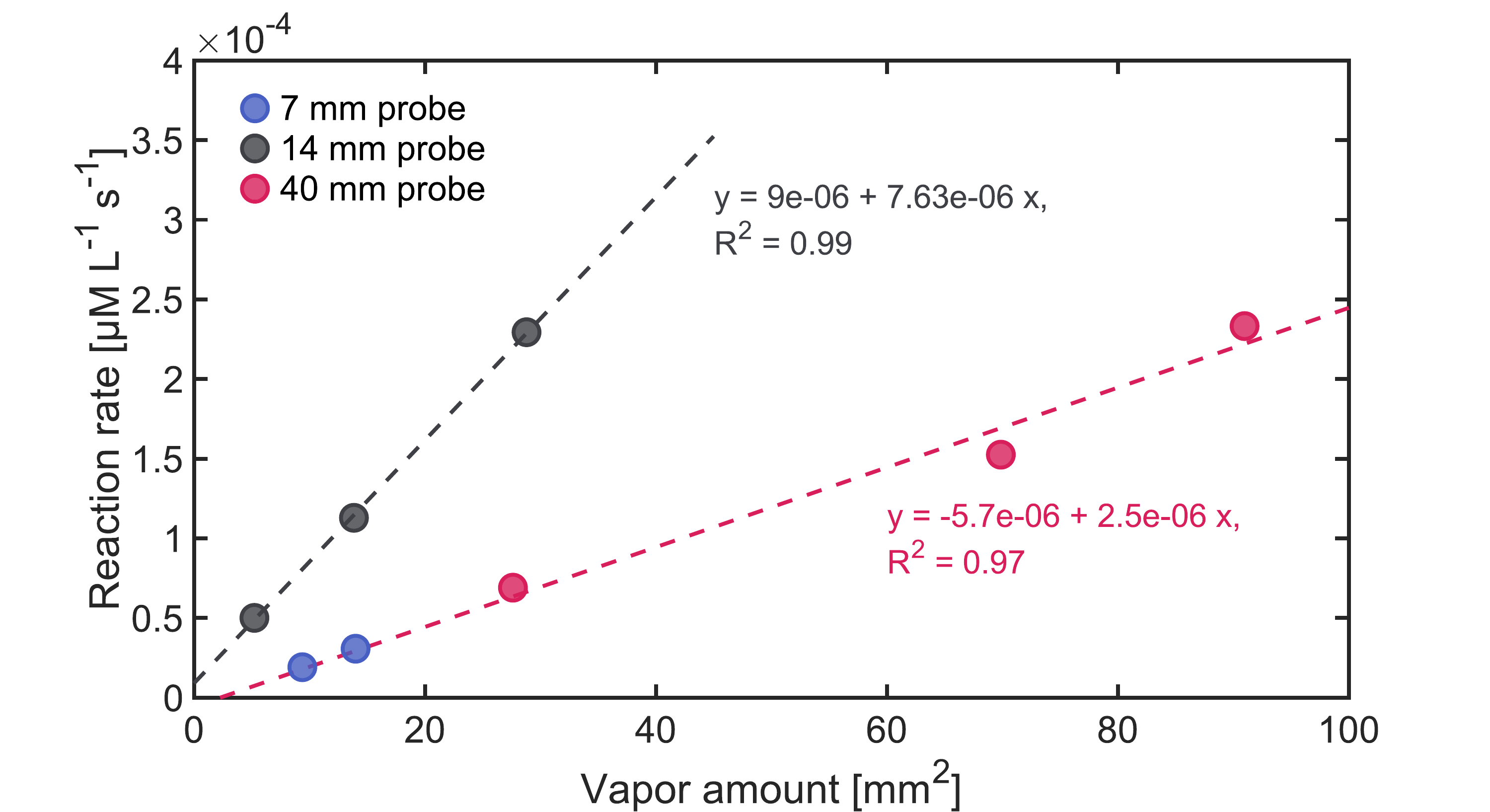}
    \caption{Initial reaction rate versus the average vapor amount. (The reaction rate was obtained by deriving the time evolution of the 7OHC concentration, Fig. \ref{Fig3}). The dashed line are the linear regressions of the scatter data.}
 
    \label{Fig7}
\end{figure*}

The analysis of the cavitation vapor field was conducted as described in section 2.3. To prove the consistency of the proposed methodology, the dependency between acoustic power, vapor amount and reaction rate was studied. 
In Fig. \ref{Fig6}, the average vapor amount is plotted against the calorimetric power for the different probe diameters.
The regressions and the coefficients of determination (Fig. \ref{Fig6}) show reasonably linear trends, with the vapor amount that tends to go to zero when the power goes to zero. For a given power, the 40 mm configuration seems to generate a higher amount of vapor compared to others. However, a direct comparison between the rates obtained from the different diameters could be inaccurate in this case. This is due to the slightly different optical setups, used for the visualization experiments, that could lead to varying sensitivities during the binarization procedure (Fig. \ref{Fig2}). Consistent trends were also obtained by plotting the reaction rate versus the average vapor amount (Fig. \ref{Fig7}). The regressions suggest a linear trend in the investigated region, with the reaction rate tending to zero when the vapor amount reaches zero.
These trends were expected since the vapor production in the domain is a consequence of the acoustic energy introduced, and the reaction rate is due to the presence of the vapor cavities, which oscillate and collapse, generating radicals. For these reasons, they prove the consistency of the adopted methodology. An alternative description of the cavitation cloud has been recently proposed by Dular and Petkov\v{s}ek, with the formulation of an acoustic cavitation parameter \citep{dular2018cavitation, kozmus2022characterization}.

Together with the total amount of vapor in the domain, the structure and the dynamics of the vapor field are also crucial to study the reactivity of the system. As observable in Fig. \ref{Fig1}, the structure of the vapor phase significantly changes from one configuration to another.

The field generated by the 3, 7 and 14 mm probe configurations is characterized by the presence of a big cluster of vapor attached to the tip of the probe. This peculiar phenomenon was first observed and studied by \v{Z}nidar\v{c}i\v{c} et al. and termed "acoustic supercavitation" \citep{vznidarvcivc2014attached,vznidarvcivc2015modeling}. 
On the contrary, for the 40 mm configuration, there is no evidence of such a cluster and the liquid is covered, more uniformly, by dispersed vapor bubbles.
To correlate the reactivity of a certain configuration with the particular structure of its vapor field, a deeper investigation of the nature of the vapor cluster (Fig. \ref{Fig1}) is required.

\begin{figure*}[t!]
    \centering
    \includegraphics[width=0.4\linewidth]{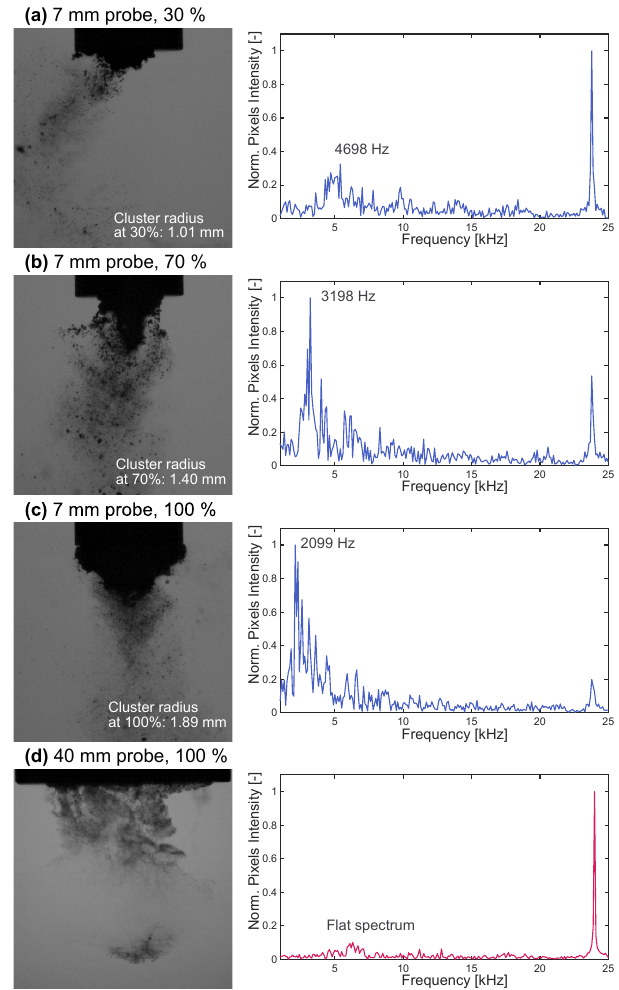}
     \caption{DFT spectra of the pixels intensity related to the near-field of the frame sequence captured with high-speed visualization.}

    \label{Fig8}
\end{figure*}

\begin{figure*}[t!]
    \centering
    \includegraphics[width=0.5\linewidth]{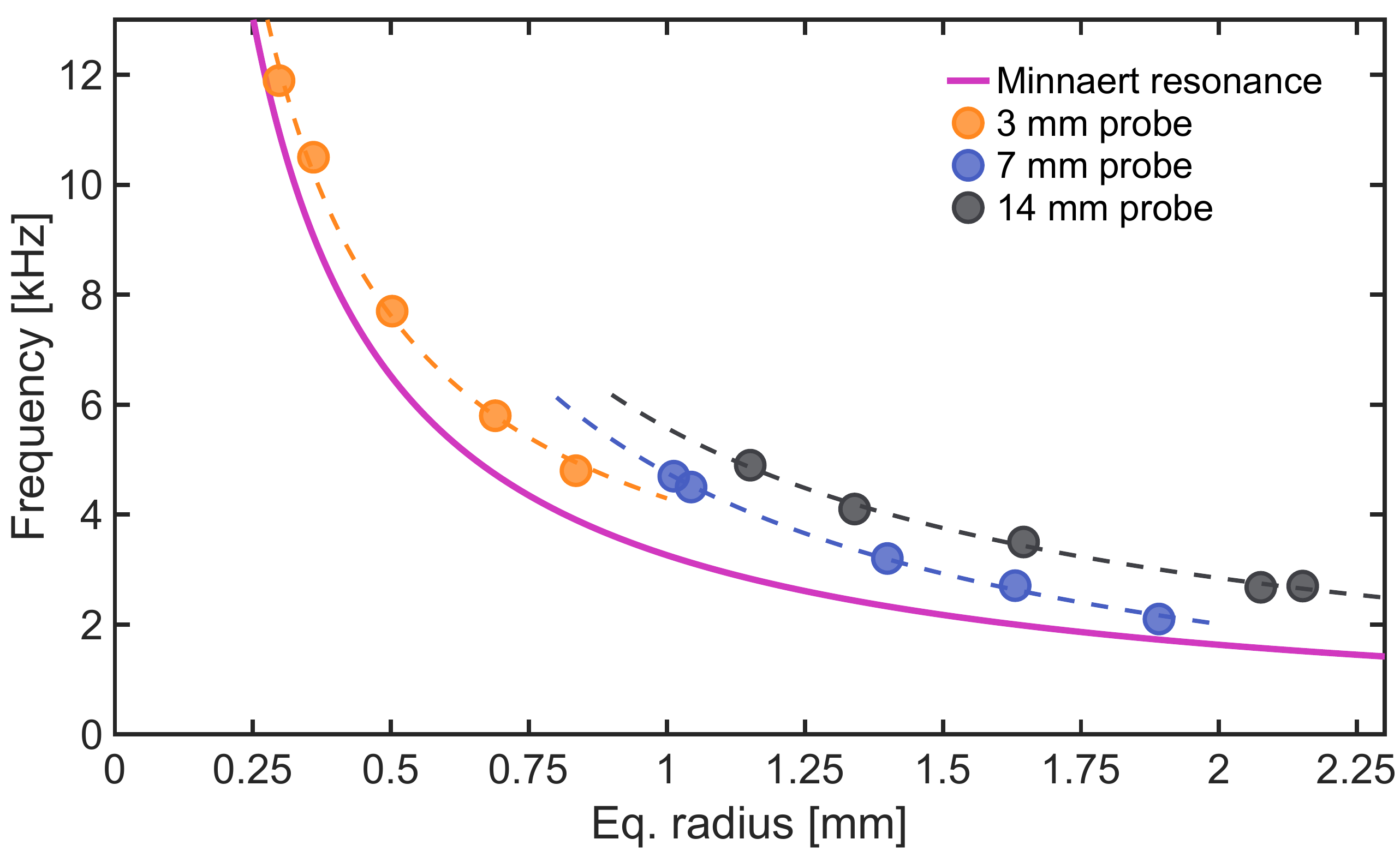}\caption{Frequency versus equivalent radius - theoretical (Minnaert resonance) and calculated (vapor cluster attached to the 3, 7 and 14 mm probes). The dashed curves represent the scatter data fitting in the form of: \(f_{r}=a/r_{eq}+b\).}
  
    \label{Fig9}
\end{figure*}

Starting from the time average of the vapor cluster area (\(A_{clust}\)), the equivalent radius can be computed using the formula: \(r_{eq}=\sqrt{A_{clust}/\pi}\).
For a given probe configuration \(r_{eq}\) can be estimated for different vibration amplitudes. In Fig. \ref{Fig8}a-c, the equivalent radius values for the 7 mm probe are indicated, together with the frequency spectrum resulting from the DFT of the pixels intensity.
In the spectra (Fig. \ref{Fig8}a, b, c), two different peaks are present: one in the low-frequency region, associated with the vapor cluster oscillation, and one at the principal frequency (probe vibration frequency: 23.8 kHz), coming from the oscillation of the bubble in the far-field of the domain. 
The different oscillation frequencies are also evident from the high-speed videos (Video 1, 2, 3 in the SI).
These results indicate that the vapor cluster oscillates at a characteristic frequency which is a decreasing function of the equivalent cluster radius. 
This trend is similar to the Minnaert resonance of a spherical bubble \citep{brennen2014cavitation, mantile2022origin}. The data obtained, for the probes where the vapor cluster is present, are plotted in Fig. \ref{Fig9}, together with the theoretical Minnaert frequency ($f_{r}=3.26/r_{eq}$). Every probe diameter exhibits a trend similar to the theoretical one. 
This confirms that the vapor cluster vibrates at its natural frequency, behaving like one single entity, like a unitary macro-bubble. The shifting of the data upwards compared to the ideal case (Minnaert frequency in Fig. \ref{Fig9}) is due to the non-spherical shape of the vapor cluster. The DFT spectrum of the pixels intensity confirms the absence of the cluster for the 40 mm probe since no relevant peaks are shown in the low-frequency range (Fig. \ref{Fig8}d). 

\subsection{Relation between working regimes and chemical efficiency}

The differences in the structure and dynamics of the vapor field are likely to influence the chemical efficiency of sonochemical systems.
In this regard, the relevant physical quantity to define two different working regimes is the acoustic intensity at the tip of the probe: \(I=\dot{W}_{in}/{S}_{t}\), where \({S}_{t}\) is the surface area of the probe tip. This quantity is listed in Tab. \ref{Tab1} for every experimental configuration. 
A high acoustic intensity characterizes the 3, 7, and 14 mm probes. For these, the high concentration of acoustic energy, near the tip of the probe, leads to the formation of the vapor cluster. After that, the acoustic energy emitted is mainly spent to sustain the oscillation of the vapor cluster, which absorbs all the radiated energy. This does not allow a uniform and deep propagation of the acoustic energy in the liquid domain, leading to a chemically inefficient system (Fig. \ref{Fig5}).

\begin{figure*}[h!]
    \centering
    \includegraphics[width=0.5\linewidth]{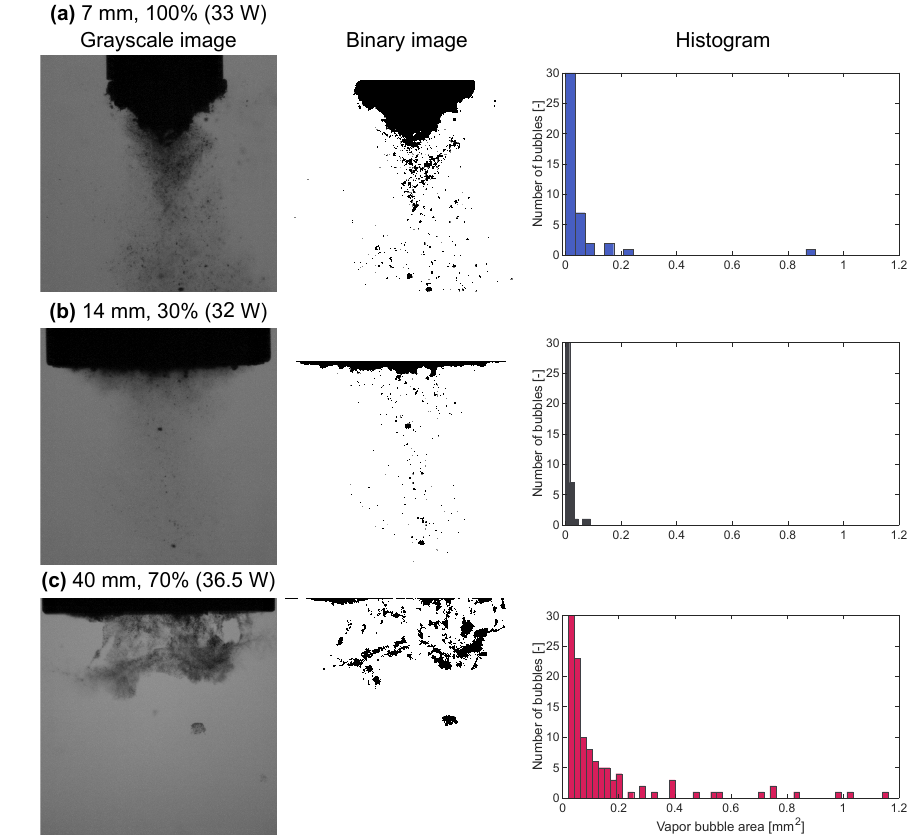}
    \caption{Histogram of the binary image for different probes diameters. In the description, between the brackets, the calorimetric power appears.}
 
    \label{Fig10}
\end{figure*}

\begin{figure*}[t!]
    \centering
    \includegraphics[width=1\linewidth]{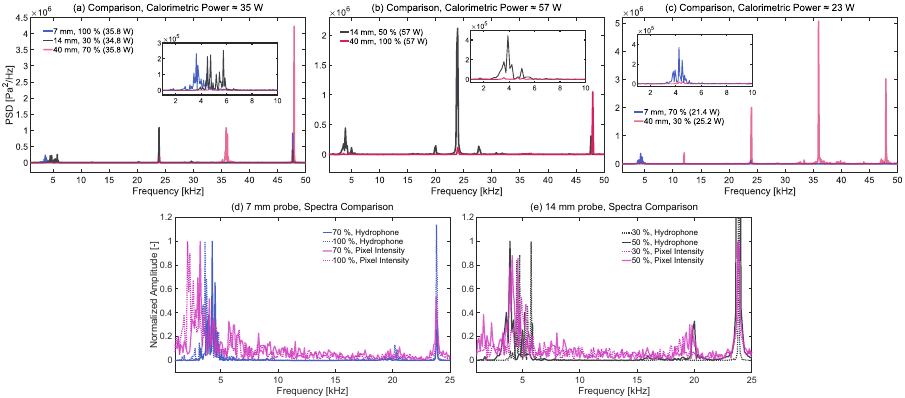}\caption{ (a, b, c) Power spectral density (PSD) of the cavitation noise, measured by the hydrophone, for different probe diameters with a similar calorimetric power.(d) Comparison of the frequency spectra coming from the hydrophone signal and pixel intensity of the high-speed sequence for the 7 mm probe. (e) Comparison of the frequency spectra coming from the hydrophone signal and pixel intensity of the high-speed sequence for the 14 mm probe.}
   
    \label{Fig9_2}
\end{figure*}

From the high-speed videos of these configurations (SI) we observe that the small bubbles, present in the far-field, "escape" from the vapor cluster and are dragged down by the secondary Bjerknes force \citep{louisnard2008analytical}.
The chemically efficient configuration (40 mm) is characterized by low acoustic intensity (Tab. \ref{Tab1}) that leads to a completely different phenomenon, where the acoustic energy density is not sufficient for the nucleation of the cluster attached to the tip. The absence of the cluster allows a more uniform and deep propagation of the acoustic waves through the domain. This triggers the nucleation of small vapor bubbles and pockets in a wider region of the liquid (Fig. \ref{Fig1}d). For the 40 mm probe, the high-speed video (Video 4 in the SI) shows that the bubbles oscillate in place at the principal frequency (23.8 kHz) and that the primary Bjerknes force is not strong enough to drag them downward. 

The differences in the structure of the vapor field, between the two regimes, are also visible in the histogram of the binary image (Fig. \ref{Fig10}). For the chemically inefficient regime (7 and 14 mm probe), the histogram shows a population of vapor bubbles having a small area (0-0.2 mm$^2$ range in Fig. \ref{Fig10}a, b). This represents the bubble cloud present in the far-field (Fig. \ref{Fig10}b, c). The vapor cluster entity, characterized by a large area, is outside the chart frame in Fig. \ref{Fig10}a, b. Instead, the 40 mm histogram (Fig. \ref{Fig10}d) shows a population of bubbles distributed over a wide range of values of the area.

In summary, the data of the coumarin conversion suggest that the highest probe diameter gave the highest efficiency in forming hydroxyl radicals (Fig. \ref{Fig5}). This is due to the particular structure of the vapor field, which is composed of a population of small vapor bubbles and pockets. As known from the literature, the production of radicals increases as the radius of the bubble decreases \citep{rivas2013ultrasound,ashokkumar2011characterization}, thanks to a higher energy density dissipated during the collapse \citep{brennen2014cavitation}. This explains the ineffectiveness in the formation of radicals of the big vapor cluster attached to the probe tip in the chemically inefficient configurations.

Additional information can be obtained by processing the acoustic signals measured by a hydrophone, immersed in the liquid domain. The frequency spectrum of an acoustic signal contains information related to the cavitation phenomena. Usually, the tonal component of the spectrum is associated with driving acoustic waves and non-inertial cavitation. At the same time, the broadband noise is a signature of the inertial or transient cavitation (bubbles collapse) \citep{mettinsound}. Since the radicals are generated only when cavitation is inertial, the energy associated with the broadband noise is expected to be significantly larger for the 40 mm configuration compared to the others.
Experiments were carried out by keeping a similar calorimetric power between the different probe configurations (Tab. \ref{Tab2}) to obtain a consistent comparison.

Starting from the raw acoustic signal (\SI{}{Pa}), registered by the hydrophone, the Power Spectral Density [PSD (\SI{}{Pa^2/Hz})] was estimated with the periodogram method \citep{MATLAB, auger1995improving, fulop2006algorithms}. The PSD spectra obtained for the different configuration comparisons are visible in Fig. \ref{Fig9_2}a, b, c.
The average power (\SI{}{Pa^2}) of the acoustic signal can be computed by integrating the PSD over a given frequency band (from 0 kHz to 600 kHz) \citep{MATLAB,auger1995improving, fulop2006algorithms}. The experimental acquisitions and the integrations were repeated three times for every configuration (Tab. \ref{Tab2}). The integral values, related to the three different acquisitions (Exp.1-3 in Tab. \ref{Tab2}), show poor repeatability. This could be due to the short time window, acquired through the oscilloscope (section 2.4), which affects the measurements. Furthermore, the average of the three measurements (Average in Tab. \ref{Tab2}) does not scale consistently with the calorimetric power provided by ultrasonic probes. The discrepancy between the calorimetric power and the PSD integral suggests that the acoustic power received by the hydrophone is just a portion of the total power introduced into the system by the probe. This is a consequence of the presence, in the domain, of the vapor phase which hinders the detection of the acoustic signal by the hydrophone. The acoustic waves, coming from different sources (probe vibration, bubble oscillation, and collapse), are reflected and scattered at the vapor/liquid interfaces. This induced the dissipation of a significant portion of the acoustic power that cannot be detected by the hydrophone.
This hypothesis is corroborated by the probe comparisons in Tab. \ref{Tab2}. With a similar calorimetry power, the PSD integral of the 40 mm configuration is significantly greater than that of the others (7 and 14 mm). This is due to the presence of the vapor cluster, in the 7 and 14 mm configuration, which reflects and dissipates the sound waves coming from the ultrasonic probe.
While, less energy is dissipated in the 40 mm configuration, where the vapor cluster is absent.

The aforementioned hydrophone detection issues do not make, a direct quantitative comparison between the PSD spectra of the different probes, possible. However, at the same time, the hydrophone signals provide qualitative information about characteristic features that differentiated the experimental configurations. The PDS spectrum of the 7 and 14 mm configurations (Fig. \ref{Fig9_2}a, b, c), exhibits a peak in the low-frequency range. This peak, which is not present in the 40 mm configurations (Fig. \ref{Fig9_2}a, b, c), represents the natural frequency of the vapor cluster oscillation and confirms the result obtained with the DFT of the pixel intensity from the high-speed imaging (Fig. \ref{Fig9}). In Fig. \ref{Fig9_2}d, e, the spectra related to the hydrophone signal and the pixel intensity are directly compared for the 7 and 14 mm probes. The two different analyses are consistent since the natural frequency peaks fall in the same frequency range. A more rigorous investigation of the noise spectra could be conducted by using the recent standard from the International Electrotechnical Commission: IEC TS 63001:2024 RLV (Measurement of cavitation noise in ultrasonic baths and ultrasonic reactors) \citep{ts2024measurement}. This was not implemented in the current study since it is not part of the primary objectives.

\begin{table}[h!]
\caption{Integral of the PSD for experimental configurations, characterized by different probe diameters and calorimetric powers (limits of integration: $a=0$ kHz, $b=600$ kHz).}
\centering
\begin{tabular}{l|r|r|r|r|r|r|r|r|r}
 & & &Exp.1 \(\int_{a}^{b} PSD \,df\)&Exp.2 \(\int_{a}^{b} PSD \,df\)&Exp.3 \(\int_{a}^{b} PSD \,df\)& Average \(\int_{a}^{b} PSD \,df\)\\
\(d\) [mm] & $A_{nom}$ [\%] &\(\dot{W}_{cal}\) [W] &  [\SI{}{Pa^2}] \(\times10^{8}\)& [\SI{}{Pa^2}] \(\times10^{8}\)& [\SI{}{Pa^2}] \(\times10^{8}\) & [\SI{}{Pa^2}] \(\times10^{8}\)\\\hline
7& 100 &35.8&3.70&3.70&3.00&3.47\\
14& 30 &34.8&3.67&2.66&2.04&2.79\\
40& 70 &35.1&36.48&36.48&13.30&28.75\\\hline
14& 50 &57.0&6.77&14.93&13.05&11.58\\
40& 100 &57.0&35.04&3.38&4.65&14.36\\\hline
7& 70 &21.4&1.90&1.99&2.44&2.11\\
40& 30 &25.2&22.59&40.92&45.06&36.19\\
\end{tabular}
\label{Tab2}
\end{table}

\section*{Conclusions}
This study investigates the relationship between radical production and the vapor field resulting from ultrasonically-induced cavitation.
The method proposed is based on coumarin dosimetry for estimating the hydroxyl radicals and high-speed imaging for the visualization of the cavitation field. The study confirms that chemical efficiency strongly depends on the structure and dynamics of the vapor field. For this reason, the probe diameter and the acoustic intensity play a crucial role. Between the four configurations tested, it was observed that the probe with the largest tip (40 mm) produced the most intense chemical activity at a given input power. It was also demonstrated that the rate of radical formation is a function of the acoustic power introduced in the domain by the ultrasonic device and linearly scales with the vapor amount.
The approach, presented in this work, can be used to characterize and optimize ultrasonic reactors working at different frequencies and for various applications involving physical or chemical treatments.

\section*{Acknowledgments}
The authors would like to thank the Core Labs, KAUST for the use of their fluorescence spectroscopy.

\bibliographystyle{unsrtnat}
\bibliography{pnas-sample}  
\end{document}